\documentclass[aps,twocolumn,nofootinbib]{revtex4-1}
\usepackage{amssymb}
\usepackage{amsmath}
\usepackage{amstext}
\usepackage{amsfonts}
\usepackage{bbold}
\usepackage{bm}
\usepackage{graphics}
\usepackage{mathtools}
\usepackage{lmodern}
\usepackage{graphicx}
\usepackage{caption}
\usepackage{subcaption}
\usepackage{mathrsfs}
\usepackage[scr=boondoxo]{mathalfa}

\newcommand{\del}{\nabla}
\newcommand{\Vtr}{V_\text{tr}}
\newcommand{\xp}{\vec x_\perp}

\newcommand{\X}{\vec X}
\newcommand{\Rp}{R_\perp}

\newcommand{\sfrac}[2]{{\textstyle\frac{#1}{#2}}}

\graphicspath{ {./} }

\begin{document}

\title{Vortex precession in trapped superfluids from effective field theory} 

\author{Angelo Esposito}
\affiliation{Center for Theoretical Physics and Department of Physics, Columbia University, New York, NY, 10027, USA}

\author{Rafael Krichevsky}
\affiliation{Center for Theoretical Physics and Department of Physics, Columbia University, New York, NY, 10027, USA}

\author{Alberto Nicolis}
\affiliation{Center for Theoretical Physics and Department of Physics, Columbia University, New York, NY, 10027, USA}

\begin{abstract}
We apply a recently developed effective string theory for vortex lines to the case of two-dimensional trapped superfluids. We do not assume a perturbative microscopic description for the superfluid, but only a gradient expansion for the long-distance hydrodynamical description and for the  trapping potential. For any regular trapping potential, we compute the spatial dependence of the superfluid density and the orbital frequency and trajectory of an off-center vortex. Our results are fully relativistic, and in the non-relativistic limit reduce to known results based on the Gross-Pitaevskii model. In our formalism, the leading effect in the non-relativistic limit arises from two simple Feynman diagrams in which the vortex interacts with the trapping potential through the exchange of hydrodynamical modes.
\end{abstract} 
\maketitle


\section{Introduction}

Vortex lines in superfluids are topological string-like objects with quantized circulation and a microscopic thickness given by the superfluid healing length. They are the only degrees of freedom that can carry vorticity, and the velocity field far from their core is irrotational but non-trivial---for a textbook treatment see~\cite{pethick2002bose,leggett2006quantum}. Indirect evidence for the presence of these objects was obtained over half a century ago in superfluid helium~\cite{Hall:1956aa,Vinen:1961aa}, and was later followed by direct observations~\cite{Williams:1974aa,Yarmchuk:1979aa,Matthews:1999aa,Chevy:2000aa,Madison:2000aa,madison3,Freilich:2010aa,Serafini:2016aa}.

\begin{figure}[b]
\centering
\includegraphics[width=0.5\textwidth]{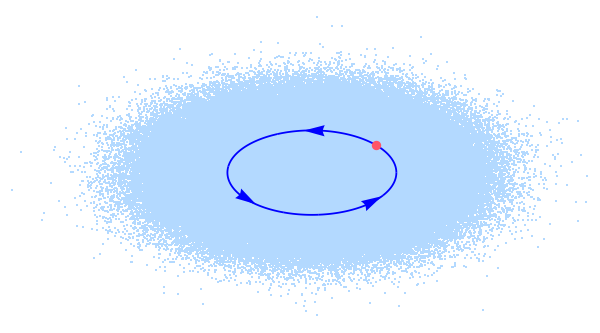}
\caption{Schematic representation of the precession of a  vortex (red dot) in an elliptical cloud.} \label{fig:cartoon}
\end{figure}

In this paper we focus on the peculiar behavior of an off-axis vortex in a non-rotating, two-dimensional trapped superfluid. The vortex is observed to orbit the center of the trapped particle cloud. If the cloud is circular, so is the orbit. If the cloud is elliptical, so is the orbit, with the same aspect ratio, as in Fig.~\ref{fig:cartoon}. For vortices close to the cloud's center, the orbital frequency is independent of the orbit's size.

The first observation of this phenomenon (known as {\it precession}) was achieved in~\cite{Anderson:2000aa} using a nearly spherical condensate of $^{87}$Rb containing a superposition of two internal components. More recently, the authors of~\cite{Ku:2014aa} performed a careful analysis of the motion and precession frequency of a vortex, using a superposition of the two lowest states of $^6$Li confined in an elliptical cloud. See also~\cite{Serafini:2015aa} for a similar study. There has already been extensive theoretical study of precession~\cite{Jackson:1999aa,fetter1,lundh,fetter2,McGee:2001aa,fetter3,Tempere2000471} in the regime where the Gross-Pitaevskii equation holds, and the solution is often treated numerically.

Here we show that the same problem can be successfully tackled using the effective field theory (EFT) methods first introduced in~\cite{Lund:1976ze,Endlich:2013dma} and recently generalized and developed in~\cite{Horn:2015zna} (see also \cite{Gubser:2014yma}). 
The main idea behind EFT is to focus directly on the long distance/low frequency degrees of freedom of a given system: one parametrizes their dynamics in terms of the most general effective action or Hamiltonian compatible with the symmetries, organized as a power expansion in energy and momentum, that is, time derivatives and spatial gradients. One of the advantages of this approach is that the validity of perturbation theory relies only on the long distance/low frequency degrees of freedom being weakly coupled, regardless of how weakly coupled the microscopic constituents of the system are. Another advantage is that one is able to derive universal predictions that follow purely from the symmetries and are insensitive to the system's microscopic physics, which might be unknown or strongly coupled, and hence intractable.

In our particular case, we will write our EFT for a generic {\em relativistic} superfluid, because we find it easier to impose the relevant spacetime symmetries in the relativistic case (see also a related discussion in \cite{Nicolis:2015sra}). However, taking the non-relativistic (NR) limit will be a simple matter of neglecting certain terms in the action explicitly suppressed by powers of $c$ (speed of light), and in that limit our derivation of the precession effect will simplify. Of course, this approach also allows us to compute the relativistic corrections to the non-relativistic result, which in principle could be important for, {\it e.g.}, neutron star physics.

In our EFT language, the phenomenon of vortex precession is entirely due to long-distance/low-energy physics and is insensitive to the microscopic physics of the superfluid or vortex core. The vortex interacts with the superfluid hydrodynamical modes, which in turn interact with the trapping potential. This leads to an indirect interaction between the vortex and the trapping potential. The symmetries of the system are so powerful that they constrain the structure of this interaction up to a few free macroscopic parameters, which can be measured experimentally.

 \vspace{1em}

\noindent {\it Conventions:} Throughout the paper we will set $\hbar=1$. We will start with $c=1$ as well, but we will later reinstate explicit factors of $c$  in order to make the non-relativistic limit straightforward and the comparison with data easier. 
We will adopt a metric signature $\eta_{\mu\nu}=\text{diag}(-,+,+,+)$. The indices $\mu,\nu,\dots$ run over all space-time coordinates,  $i,j,\dots$ run over spatial coordinates only, and $\alpha, \beta, \dots$ run over worldsheet coordinates. 


\section{The untrapped action} \label{review}

We now review the main aspects of the effective theory presented in~\cite{Horn:2015zna}, which will require some familiarity with high energy ideas. The reader who is unfamiliar or uninterested in these concepts may refer directly to Eqs.~\eqref{eq:S2}--\eqref{couplings} below, where we report the simple action for the interaction of superfluid modes with vortex lines.

The EFT description of a superfluid with vortices involves a two-form field ${\cal A}_{\mu\nu}(x)$ for the bulk degrees of freedom and an embedding position field $X^\mu(\tau, \sigma)$ for each vortex, where $\tau$ and $\sigma$ are arbitrary worldsheet coordinates, such as proper time and physical length along the vortex. To lowest order in derivatives and in the case of a single vortex, the action reads \cite{Horn:2015zna} 
\begin{align}
S & = S_\text{bulk} + S_\text{KR} + S_{\text{NG}^\prime} + \mbox{gauge fixing}\\
S_\text{bulk} & \equiv \int d^4x\,G(Y) \; , \qquad Y=-F_\mu F^\mu \\
S_\text{KR} &\equiv \lambda \int d\tau d\sigma\, {\cal A}_{\mu\nu}\partial_
\tau X^\mu\partial_\sigma X^\nu\; \label{eq:KR} \\
S_{\text{NG}^\prime}& \equiv-\int d\tau d\sigma\,\sqrt{-\text{det}\,g}\,\mathcal{T}\left(g^{\alpha\beta}h_{\alpha\beta},Y\right)  \label{eq:NG} \; .
\end{align}
Here $F^\mu=\frac{1}{2}\epsilon^{\mu\nu\rho\sigma}\partial_\nu {\cal A}_{\rho\sigma}$ is the gauge invariant field strength for ${\cal A}$, $G$ is an {\it a priori} arbitrary function, $\lambda$ is a coupling constant, ${\cal T}$ is a generalized tension, and 
\begin{align}
g_{\alpha\beta}=\eta_{\mu\nu}\partial_\alpha X^\mu\partial_\beta X^\nu \; , \quad h_{\alpha\beta}=\frac{F_\mu F_\nu}{Y}\partial_\alpha X^\mu\partial_\beta X^\nu
\end{align}
are two independent induced worldsheet metrics. The local values of the superfluid number density $n$ and (relativistic) energy density $\rho$ are related to $Y$ and $G$ by
\begin{align}
Y = n^2 \; , \qquad G(Y) = - \rho \; ,
\end{align}
so that the superfluid equation of state $\rho = \rho(n)$ uniquely determines the bulk action $G(Y)$.

A homogeneous superfluid at rest with number density $\bar n$ corresponds to a background field ${\cal A}_{ij} = -\frac13 \bar n \, \epsilon_{ijk} x^k$. If one is interested in studying superfluid configurations close to such a state, one can expand the above action in powers of perturbations of ${\cal A}$ about its background and apply standard perturbative field theory techniques. Since at some point we will be taking the NR limit, it is useful to be explicit about powers of $c$. Hence, we parametrize the perturbations $\vec A$ and $\vec B$ of ${\cal A}_{\mu\nu}$ as
\begin{align}
{\cal A}_{0i}={\bar n}A_i(x)/c \; , \quad {\cal A}_{ij}=\bar n \, \epsilon_{ijk}\left(-\sfrac{1}{3}x^k+B^k(x)\right).
\end{align}
In this way, they both have regular propagators in the $c\to \infty$ limit. Indeed, choosing the $\tau = t =X^0/c$ gauge, the expansion of the action above reads \cite{Horn:2015zna}
\begin{align} \label{eq:S2}
S& \to \frac{\bar w}{c^2}\!\int \!d^3\!x dt\Big[\sfrac{1}{2}\big(\vec\del\times\vec A \, \big)^2+\sfrac{1}{2}\big(\dot{\vec B}^2-c_s^2(\vec\del\cdot\vec B)^2\big) \\
& \qquad\;\;+\sfrac{1}{2}\left(1-\sfrac{c_s^2}{c^2}\right)\vec\del\cdot\vec B\big(\dot{\vec B}-\vec\del\times\vec A\big)^2\Big] \notag \\
& -\int dtd\sigma \Big[\sfrac{1}{3} {\bar n\lambda} \, \epsilon_{ijk}X^k\partial_tX^i\partial_\sigma X^j + T_{(00)} \big|\partial_\sigma\vec X\big| \Big] \label{free string} \\
&+ \int dtd \sigma \,  \Big[ {\bar n\lambda} \big(A_i\partial_\sigma X^i + \epsilon_{ijk}B^k\partial_tX^i\partial_\sigma X^j \big) \label{couplings}\\
&\quad + \big|\partial_\sigma\vec X\big|\Big(2T_{(01)}\vec\del\cdot\vec B+2T_{(10)}\big(\dot{\vec B}-\vec \del\times\vec A\big)\cdot \frac{\vec v_\perp}{c^2}\Big) \Big]\; , \notag
\end{align}
where $\bar w$ is the background relativistic enthalpy density ($\simeq \mbox{mass density} \times c^2$, in the NR limit), $c_s$ is the sound speed,  the $T$s are effective couplings obtained from the generalized string tension by
\begin{align}
T_{(mn)}= a^m b^n\frac{\partial^m}{\partial a^m}\frac{\partial^n}{\partial b^n}\mathcal{T}(a,b) \; ,
\end{align}
evaluated on the background, and $\vec v_\perp$ is the local string's velocity in the direction orthogonal to the string itself. For non-relativistic superfluids, the constant $\lambda$ is related to the vortex's circulation $\Gamma$ by $\lambda = m \Gamma$, where $m$ is the mass of the superfluid's microscopic constituents. We stopped the expansion at cubic order in the bulk and at linear order on the worldsheet, since we will not need higher order terms. The cubic term that we have kept (second line of Eq.~\eqref{eq:S2}) is  known~ to play a role in the classical running of $T_{(01)}$ \cite{Horn:2015zna}, and it will be important for us as well. We also implicitly chose a gauge fixing term for ${\cal A}_{\mu\nu}$ that makes $\vec A$ purely transverse and $\vec B$ purely longitudinal---again, see \cite{Horn:2015zna}. $\vec B$ can thus be identified with the phonon field, while $\vec A$ is a constrained field playing a role similar to that of the Coulomb potential of electrodynamics: it does not feature propagating wave solutions, but it can mediate long-range interactions between sources (vortex lines, in this case). It has been dubbed the {\it hydrophoton} \cite{Endlich:2013dma}.

For convenience for what follows, we organized the expanded action in this way: the first integral (Eq.~\eqref{eq:S2}) collects the terms that make up the action for the bulk $\vec A$ and $\vec B$ fields, {\it i.e.} the action describing the superfluid in the absence of vortices. The associated propagators are
\begin{align} \label{eq:G}
G^{ij}_A(k)=\frac{c^2}{\bar w}\frac{i (\delta^{ij}-\hat k^i\hat k^j)}{k^2}, \;\, G_B^{ij}(k)=\frac{c^2}{\bar w}\frac{i \,\hat k^i\hat k^j}{\omega^2-c_s^2k^2},
\end{align}
where  the $i\epsilon$ prescription is understood. The second integral (Eq.~\eqref{free string}) describes the motion of a free vortex line in an unperturbed superfluid. The last integral (Eq.~\eqref{couplings}) collects all the interaction terms between the bulk fields and the string. In the units that we are using, $\lambda$ is dimensionless, all the $T$s have units of tension (energy per unit length), $\bar n$ is a number density, and $\bar w/c^2$ is a mass density. All these coupling constants are finite in the $c \to \infty$ limit, so the only suppressed term is the last one, which involves an explicit $1/c^2$ factor.


\section{Modeling trapping} \label{density}

Now that we have set up the formalism, we can perform our analysis. Let us forget for the moment about the presence of the vortex and focus on the superfluid only. 
To describe the spatial confinement of the superfluid, we introduce a position-dependent action term for the superfluid modes. 
In line with the EFT approach, we write the most general trapping term that, at lowest order in derivatives, is compatible with the symmetries of our system:
\begin{align} \label{eq:Str}
S_\text{tr}=-\int d^3 x dt \,{\cal E} \!\big(\sqrt{Y},\vec u,\vec x \big)\; ,
\end{align}
where
\begin{align}
\vec u=\frac{\dot{\vec B}-\vec\nabla\times\vec A}{1-\vec\nabla\cdot\vec B}
\end{align}
is the superfluid velocity field, and for now ${\cal E}(\sqrt{Y},\vec u,\vec x)$ is a generic function of its arguments, with units of energy density.

Noting that $Y=\bar n^2\big[(1-\vec\del\cdot\vec B)^2-\frac{1}{c^2}(\dot{\vec B}-\vec\del\times\vec A)^2 \big]$ and expanding in perturbations of $Y$ and $\vec u$, we get new interaction terms for $\vec A$ and $\vec B$ of the form
\begin{align}
S_{\rm tr} &\to   \int d^3x dt \,  \Big\{ \bar n V(\vec x)  \Big[\vec\del\cdot\vec B +
\sfrac{1}{2c^2}\big( \dot{\vec B}-\vec\del\times\vec A\,\big)^2 \Big] \notag  \\
&- \sfrac{1}{2}\rho_{ij}(\vec x)\big(\dot{\vec B}-\vec\nabla\times\vec A\,\big)^i\big(\dot{\vec B}-\vec\nabla\times\vec A\,\big)^j \Big\} \label{eq:SI} 
\end{align}
where 
\begin{align}
V(\vec x) \equiv \frac{\partial {\cal E}}{\partial \sqrt{Y}}  \; , \qquad \rho_{ij}(\vec x) \equiv \frac{\partial^2 {\cal E}}{\partial u^i \partial u^j} \; ,
\end{align}
both evaluated on the background ($\sqrt Y = \bar n$, $\vec u = 0$). Notice that $V$ has units of energy, and $\rho_{ij}$ has units of mass density. We are assuming that the trapping mechanism does not involve a breaking of time reversal, and we are thus setting to zero terms with odd powers of $\vec u$. 
If this assumption is violated---say, as in the case of magnetic trapping of charged particles---then the second line in Eq.~\eqref{eq:SI} should be replaced with a linear term in $\vec u$. 
Lastly, we have kept only the lowest order terms for any combination of derivatives (time or space) and fields ($A$ or $B$).  This truncation is all we need to compute lowest-order results in perturbation theory.

In the standard Gross-Pitaevskii approach, the confinement of the superfluid is modeled with an interaction between a trapping potential $\Vtr(\vec x)$ and the superfluid density only. This amounts to considering the particular case of ${\cal E}(Y,\vec u,\vec x)=\Vtr( \vec x)\sqrt{Y}$. Our  trapping action \eqref{eq:Str} is a more general starting point. Notice, however, that to lowest order in perturbation theory and if we neglect the velocity dependence of ${\cal E}$, the two approaches coincide.

To first order in $V$ and $\rho_{ij}$, the action~\eqref{eq:SI} provides an external source for the $\vec B$ field. This is given by
\begin{align} \label{eq:JBV}
\vec J_B(x)=-\bar n\,\vec \nabla V(\vec x) \; .
\end{align}
From standard Green's functions theory, the expectation value for $\vec B$ in the presence of a source is
\begin{align}
\langle B^i(x) \rangle =\int\frac{d^3k d\omega}{(2\pi)^4}iG_B^{ij}(k)J_B^j(k)\,e^{i k \cdot x} \; ,
\end{align}
and one easily finds
\begin{align}
\langle \vec\del\cdot\vec B \rangle=\frac{\bar n c^2}{\bar w c_s^2}V(\vec x) \; .
\end{align}
Since there are no linear sources for $\vec A$, this implies that the superfluid density in the presence of (weak) trapping is
\begin{align} \label{eq:nx}
n(\vec x)= \sqrt{Y} = \bar n\left(1-\frac{\bar n c^2}{\bar w c_s^2}V(\vec x)\right) \; .
\end{align}
It is interesting to note that, to this order in perturbation theory, the geometry of the density field is the same as that of the trapping potential, in the sense that the two have the same level surfaces. This is due to the derivatives entering $\vec J_B$ and  $Y$: they compensate for the non-locality of the propagator, effectively turning the interaction with the trap into a contact term. 

Our result in Eq.~\eqref{eq:nx} is completely general, valid for any relativistic superfluid. In the non-relativistic limit one has $\bar w\simeq m\bar n c^2$, where $m$ is the mass of the microscopic constituents of the superfluid, and the expression above simplifies to
\begin{align} \label{eq:nxNR}
n(\vec x) \to \bar n \left(1-\frac{V(\vec x)}{m c_s^2} \right) \; ,
\end{align}
which matches the standard result obtained in the Thomas-Fermi approximation---see {\it e.g.}~\cite{pethick2002bose}.\footnote{In the Gross-Pitaevskii approximation one also has $c_s^2=\mu/m$ and $\mu=nU_0$, where $\mu$ is the chemical potential and $U_0$ is the coupling of the non-linear interaction.}

\begin{figure} [t]
   \centering 
   \includegraphics[width=0.45\textwidth]{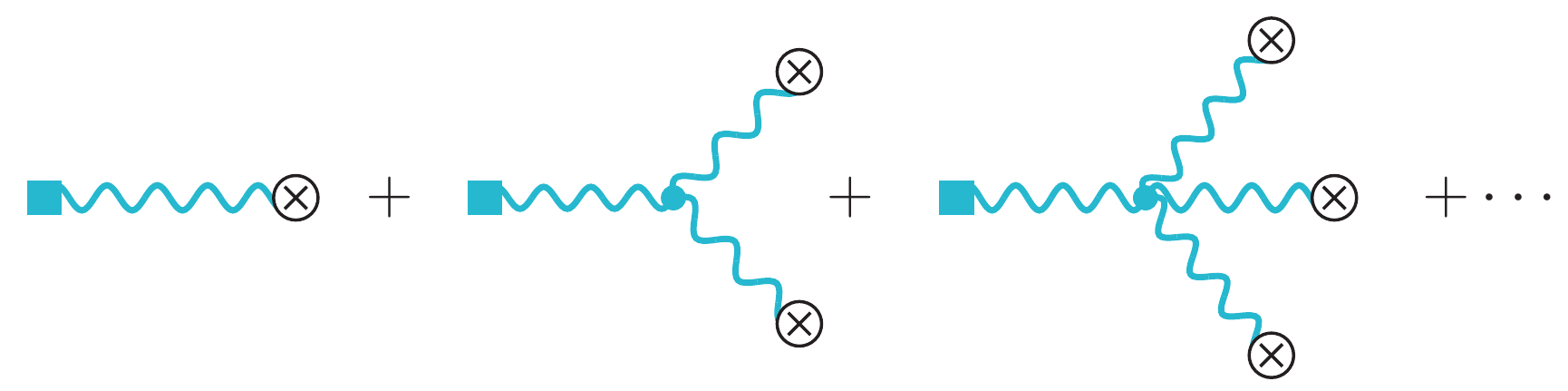}
   \caption{Non-linear corrections to Eqs.~\eqref{eq:nx}, \eqref{eq:nxNR}. The crossed circles depict the trapping potential $V(\vec x)$, the wavy lines propagators of $\vec B$, the dots self-interactions of $\vec B$, and the squares the density $n(\vec x)$.
   }
   \label{resum}
\end{figure}

Since we have been working to first order in the trapping potential, Eqs.~\eqref{eq:nx} and \eqref{eq:nxNR} can be trusted only when $V(\vec x)$ can be treated as a small perturbation, which is certainly not the case close to the edge of the cloud. However, the spatial point about which we are expanding is arbitrary, and so our expansion is really an expansion in small {\em variations} of $V$, or, equivalently, small gradients (in units of the healing length). Following standard renormalization group (RG) logic, we can thus rewrite Eqs.~\eqref{eq:nx} and \eqref{eq:nxNR} as differential equations, the non-linear solutions of which are valid to all orders in $V$. They formally correspond to resumming  an infinite series of tree-level diagrams, which for the case at hand are those of Fig.~\ref{resum}.
In particular, in the non-relativistic case one can rewrite Eq.~\eqref{eq:nxNR} as
\begin{equation} \label{RG n}
c_s^2(n) \frac{dn}{n} = - \frac{dV}{m}  \; .
\end{equation}
If one knows the equation of state of the superfluid, and thus $c_s^2(n)$, one can integrate both sides and find the fully non-linear relationship between $n(\vec x)$ and $V(\vec  x)$. For instance, in Gross-Pitaevskii theory one has $c_s^2 \propto n$, and thus the linearized solution \eqref{eq:nxNR} is in fact valid to all orders in $V$, in agreement with standard results. For more general (and realistic) equations of state, there can be sizable nonlinear corrections.


\section{Vortex Precession in two dimensions} \label{2D}

We will now study the motion of a single vortex in a trapped superfluid. In the field theoretical approach, this is done by integrating out the superfluid's bulk modes (the $A$ and $B$ fields). This results in an effective interaction between the vortex line and the trapping potential, which,
in our formalism, is at the origin of the observed precession of the vortex.

For simplicity we consider the cylindrical case only, that is, a three-dimensional superfluid trapped only along the $(x,y) \equiv \vec x_\perp$ directions. We parametrize our vortex as a straight line, $\X(t,z)\equiv\left(X(t),Y(t),z\right)$, and we assume that its distance from the center is much smaller than the typical transverse size of the cloud. 
Moreover,  we will work in the non-relativistic limit, which is the relevant one for experimental questions. (The EFT language allows one to compute relativistic corrections with little extra work; we present such corrections in the Appendix.) To this end, we will assume that the second line of Eq.~\eqref{eq:SI} is a relativistic correction, {\it i.e.} it is secretly suppressed by inverse powers of $c$, since it describes a direct coupling of the trap to the superfluid velocity; 
this can come, for example, from Doppler-like effects~\cite{privatecom}, which are indeed suppressed by inverse powers of $c$.  Thus, we assume
\begin{equation} \label{rhoij}
\rho_{ij} (\vec x ) = \sfrac{1}{c^2} \bar n V_{ij}(\vec x) \; , \qquad V_{ij} \sim V \; .
\end{equation}
We emphasize, however, that for magnetic trapping of charged particles this assumption should be lifted, and in fact the second line in Eq.~\eqref{eq:SI} should be replaced by a linear coupling to $\vec u$.

So, in the $c\to \infty$ limit, the only surviving interaction terms for $A$ and $B$ in Eqs.~\eqref{eq:S2}, \eqref{couplings} and \eqref{eq:SI} are the cubic vertex as well as the sources
\begin{align}
\label{eq:JA} \vec J_A(x) =\bar n & \lambda  \delta^2(\xp-\X) \hat z \\
\label{eq:JB} \vec J_B(x) = \big[ & \left(\bar n\lambda\epsilon_{ab}\dot X^b - 2 T_{(01)}\partial_a\right)\delta^2(\xp-\X) \nonumber \\
& -\bar n \partial_aV(\xp)\big]\hat x_\perp^a \; , 
\end{align}
where from now on the indices $a,b$ run over $x,y$.

\begin{figure} [t]
   \centering 
   \includegraphics[width=0.4\textwidth]{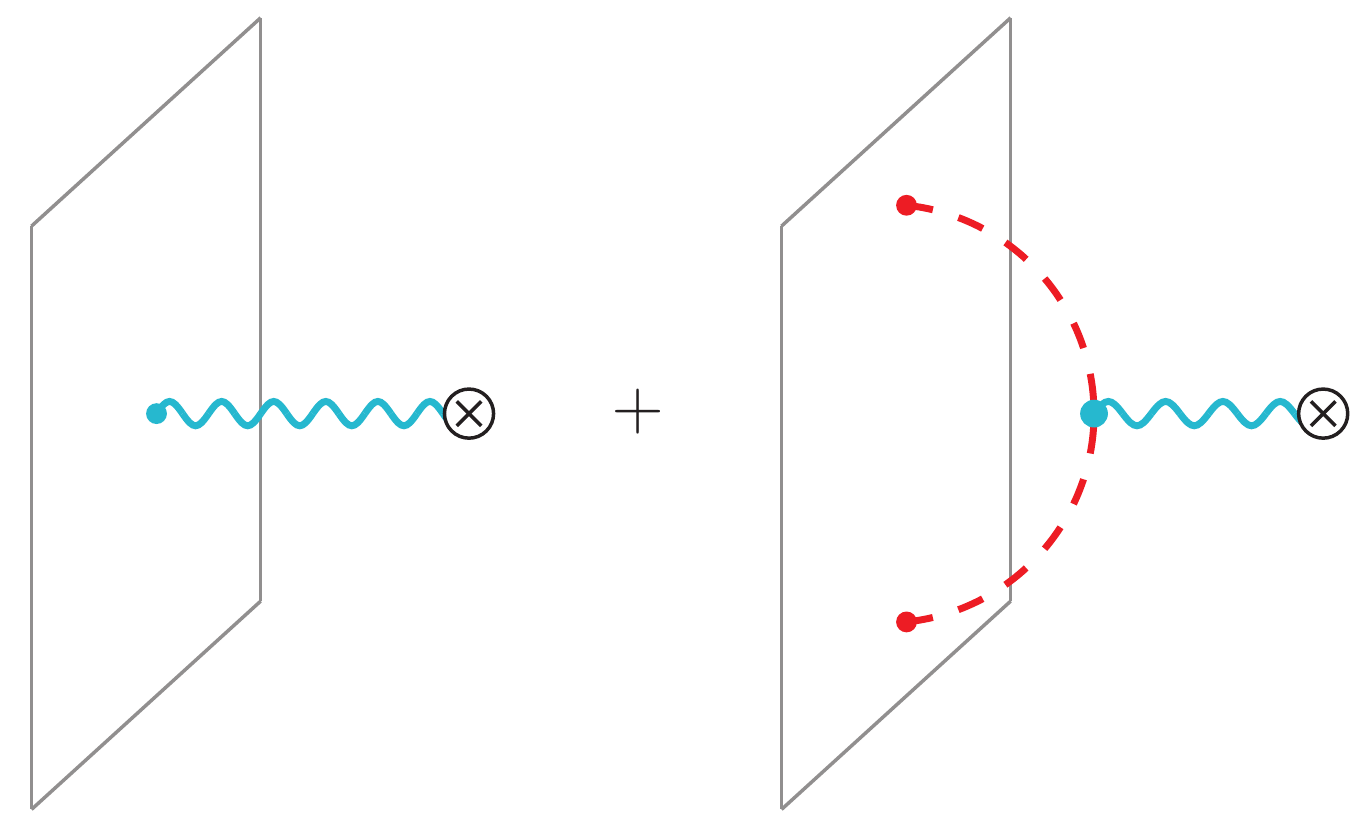}
   \caption{Feynman diagrams representing the non-relativistic interactions between the vortex worldsheet (plane) and the trapping potential (crossed circle), mediated by phonons (wavy blue line) or hydrophotons (dashed red line). 
   }
   \label{fig:diagrams}
\end{figure}
Now consider integrating out $A $ and $B$ along the lines of the field theoretical methods of~\cite{Horn:2015zna}. At tree level this amounts to replacing them in Eq.~\eqref{eq:S2} with the solutions to their classical equations of motion. The corresponding corrections to the vortex effective action are of the form
\begin{align} \label{eq:SeffAB}
S_\text{eff}^{(A)}[\vec X]&=\int \frac{d^3kd\omega}{(2\pi)^4}J_A^i(-k)iG_A^{ij}(k)J_A^j(k)  \\
S_\text{eff}^{(B)}[\vec X ]&=\int  \frac{d^3kd\omega}{(2\pi)^4}J_B^i(-k)iG_B^{ij}(k)J_B^j(k) \; . \label{eq:SeffB}
\end{align}
These will both have mixed terms that make the vortex interact with the trapping potential, corresponding to the diagrams in Fig.~\ref{fig:diagrams}. To compute these, we can neglect the time-derivative term in $\vec J_B$, since there is already a one-derivative kinetic term for the string in Eq.~\eqref{free string}, and thus any 
${\cal O}(V)$ corrections to it can be neglected in first approximation. On the other hand, there is no position-dependent potential for the string in the free string action---the only breaking of translational invariance comes from the trapping potential---so whatever position-dependent ${\cal O}(V)$ non-derivative term we get will be the leading source of ``forces" for the string.

In conclusion, we can simply restrict to the (Fourier-space) sources
\begin{align} \label{eq:JBk}
\vec J_A(k)\!&=\! \hat z\,\bar n\lambda(2\pi)^2 \delta(\omega)\delta(k_z)e^{-i\vec{k}_\perp\cdot\X} \; ,\\
\vec J_B(k)\!&=\!- i(2\pi)^2\delta(\omega)\delta(k_3)\vec k_\perp\left[ 2T_{(01)}e^{-i\vec{k}_{\perp}\!\cdot\X}+\bar nV(\vec{k}_{\perp})\right].
\end{align}
If the source for $B$ is substituted into Eq.~\eqref{eq:SeffB}, it produces the following contribution:
\begin{align} \label{eq:SB}
   S_{\text{eff}}^{(B)} [\vec X]  \supset \frac{2T_{(01)}}{m c_s^2}\int dtdz\,V(\X) \; .
\end{align}
(We used the fact that for a non-relativistic superfluid $\bar w \simeq \bar n  m c^2 $).
Once again, the derivatives entering the interactions of $\vec B$ cancel the non-locality of its propagator, and the net result is a purely local interaction between the vortex and the trapping potential\footnote{This is {\em not} a general phenomenon: when integrating out gapless modes, one in general expects to get long-range interactions.}.

The cubic interaction $(\vec{\nabla}\cdot\vec{B})(\vec{\nabla}\times\vec{A})^2$ of Eq.~\eqref{eq:S2} in the presence of a non-trivial $V$ can instead be thought of as a modification to the hydrophoton propagator (see the second diagram in Fig.~\ref{fig:diagrams}), which plugged into Eq.~\eqref{eq:SeffAB} gives
\begin{align} \label{eq:SA}
   S_{\text{eff}}^{(A)} [\vec X]  \supset \frac{\bar n^3\lambda^2c^4}{8\pi^2\bar w^2c_s^2}\left(1-\frac{c_s^2}{c^2}\right)\int dt dz d^2x_\perp \frac{V(\vec x_\perp+\vec X)}{x_\perp^2} \; .
\end{align}

Up to terms that do not depend on the vortex position, the complete NR vortex effective action is therefore
\begin{widetext}
\begin{align} \label{eq:SNR}
  S^{(\text{NR})}_{\text{eff}} [\vec X] = \!\int\! dtdz\left[\frac{\bar n\lambda}{3}\epsilon_{ab}X^a\dot X^b+\frac{2T_{(01)}}{m c_s^2}V(\X)+\frac{\bar n\lambda^2}{8\pi^2m^2c_s^2}\int d^2x_\perp\frac{V(\vec x_\perp+\vec X)}{x_\perp^2}\right]\!,
\end{align}
\end{widetext}
and studying the motion of the vortex is now reduced to a straightforward problem of point particle mechanics in two dimensions. 
Note that the last term is non-local, in the sense that it involves values of $V$ away from $\vec X$. It is effectively a long-range interaction between the trapping potential and the vortex. 

The equations of motion read
\begin{equation} \label{eom}
\frac{2 \bar n \lambda}{3} \epsilon_{ab} \dot X^b - \partial_a V_{\rm eff} (\vec X) = 0 \; ,
\end{equation}
where the vortex's effective potential energy (per unit length) is
\begin{equation} \label{V eff}
V_{\rm eff} (\vec X) \equiv - \frac{2T_{(01)}}{m c_s^2}V(\X) -\frac{\bar n\lambda^2}{8\pi^2m^2c_s^2}\int d^2x_\perp\frac{V(\vec x_\perp+\vec X)}{x_\perp^2} \; .
\end{equation}

Things become more transparent if we consider a vortex close to the center of the cloud. In that case, we can expand $V_{\rm eff}$ for small $\vec X$ and, to quadratic order, we get
\begin{align} \label{expanded Veff}
V_{\rm eff} (\vec X) & \simeq -\frac12 X^a X^b \bigg[ \frac{2T_{(01)}}{m c_s^2}\partial_a \partial_b V(0) \nonumber \\
& + \frac{\bar n\lambda^2}{8\pi^2m^2c_s^2}\int d^2x_\perp\frac{\partial_a \partial_b V(\vec x_\perp)}{x_\perp^2} \bigg].
\end{align}
(We are assuming that the linear terms vanish---that is for us what defines the ``center'' of the cloud).
There are now two qualitatively different cases, depending on whether $\partial_a \partial_b V(0)$ vanishes or not.


\subsection{Harmonic trapping ($\partial_a \partial_b V(0) \neq 0$)}

In the case of approximately harmonic trapping, $V(\vec x_\perp)$ is quadratic close to the center of the cloud, so
$\partial_a \partial_b V(0)$ does not vanish. Then the first line in Eq.~\eqref{expanded Veff} is nonzero, and the integral in the second line has a logarithmic divergence at $\vec x_\perp = 0$:
\begin{align}
\int d^2x_\perp\frac{\partial_a \partial_b V(\vec x_\perp)}{x_\perp^2} = - \partial_a \partial_b V(0) \cdot 2\pi \log a + \dots
\end{align}
where $a$ is an arbitrary UV cutoff length, and the dots stand for terms that are finite for $a \to 0$. As usual with UV log divergences, by dimensional analysis they must be accompanied by the log of a physical infrared scale. In the integral above, the only (implicit) candidate for such a scale is the typical transverse size of the cloud $R_\perp$; our perturbation theory breaks down there, thus making the extrapolation of such integrals beyond that point nonsensical.  We find
\begin{align}
\int d^2x_\perp\frac{\partial_a \partial_b V(\vec x_\perp)}{x_\perp^2} = \partial_a \partial_b V(0)\, 2\pi \log (R_\perp/ a) \; ,
\end{align}
so the second line of Eq.~\eqref{expanded Veff} can be thought of as a renormalization of the first. 

In particular, following  standard RG ideas, we can parametrize $V_{\rm eff}$ in terms of a running coupling $T_{(01)}(q)$ evaluated at a typical momentum $q \sim 1/R_\perp$:
\begin{align}
V_{\rm eff} (\vec X) & \simeq - \frac{T_{(01)}(1/R_\perp)}{m c_s^2}\partial_a \partial_b V(0) \, X^a X^b \, ,
\end{align}
where
\begin{align}
T_{(01)}(q) = - \frac{\bar n\lambda^2}{8\pi m}\log(q \ell) \; ,
\end{align}
and $\ell$ is a physical microscopic scale, which we expect to be of the order of the healing length, but the precise value of which has to be determined from experiments. Notice that this result matches precisely the running of $T_{(01)}$ found in \cite{Horn:2015zna} via somewhat different methods.

If we parametrize the harmonic trapping potential in the usual elliptical form,
\begin{align} \label{eq:ellip}
V(\xp) = \frac{m}{2}\left(\omega_x^2 x^2+\omega_y^2 y^2\right)+\mathcal{O}(r^4)\; ,
\end{align} 
the equations of motion \eqref{eom} reduce to
\begin{align} \label{eq:eom}
\dot X(t)=\omega_p\frac{\omega_y}{\omega_x} Y(t) \; ,\quad \dot Y(t)=-\omega_p\frac{\omega_x}{\omega_y} X(t) \; ,
\end{align}
the solutions of which are elliptical orbits with the same orientation and aspect ratio as the trapping potential, with angular frequency
\begin{align} \label{eq:omegap}
\omega_p \equiv \frac{3\Gamma}{8\pi c_s^2}\omega_x\omega_y \log(R_\perp/\ell)\, .
\end{align}
(We  used that $\lambda=m\Gamma$ in the NR limit.)
This matches the standard results derived by more traditional methods \cite{Jackson:1999aa,fetter1,fetter2}. Nonetheless, we emphasize the generality of our result: it does not rely on the Gross-Pitaevskii model or the Hartree approximation.

In fact, one can go further and make Eq.~\eqref{eq:omegap} completely predictive. Recall that at this level $\ell$ is a free parameter the value of which has to be determined by experiment. However, if we consider the combination
\begin{align}
\chi \equiv \omega_p/\omega_x \omega_y
\end{align}
and compare the values it takes for different trapping potentials (`1' and `2'), the $\ell$ dependence cancels out and we get
\begin{align} \label{chi}
\chi_1 - \chi_2 = \frac{3 \Gamma}{8 \pi   c_s^2} \,  \log\frac{\Rp {}_{,1}}{\Rp {}_{,2}} \; .
\end{align}


\subsection{Flatter trapping potentials ($\partial_a \partial_b V(0) = 0$)} \label{sec:shallow}

The case of flatter trapping potentials---{\it i.e.} such that $\partial_a \partial_b V(0) = 0$---is easier to study:  the first line in Eq.~\eqref{expanded Veff} is zero, and the integral in the second line is convergent at $\vec x_\perp = 0$. 

Consider then parametrizing the trapping potential as
\begin{align} \label{parametrize}
V(\vec x_\perp) = m c_s^2 \, f(\vec x_\perp/R_\perp) \; , 
\end{align}
where $R_\perp$ is again the typical transverse size of the cloud, $f$ is a dimensionless function generically with order 1 coefficients (but vanishing second derivatives at the origin), and the overall prefactor follows from consistency with Eq.~\eqref{eq:nxNR}.
For the integral in \eqref{expanded Veff}, we now simply have
\begin{align} \label{finite integral}
\int d^2x_\perp\frac{\partial_a \partial_b V(\vec x_\perp)}{x_\perp^2} = \frac{m c_s^2}{R_\perp^2} \, f_{ab} \; ,
\end{align}
where $f_{ab}$ is a constant symmetric tensor with order 1 entries.

If we align the $x$ and $y$ axes with the eigenvectors of $f_{ab}$, the equations of motion \eqref{eom} read
\begin{align} 
\dot X(t)=\omega_p \sqrt{\frac{f_{yy}}{f_{xx}}} \, Y(t) \; ,\quad \dot Y(t)=-\omega_p \sqrt{\frac{f_{xx}}{f_{yy}}} \, X(t) \; ,
\end{align}
the solutions of which now are elliptical orbits with aspect ratio $\sqrt{f_{yy}/f_{xx}}$ and angular frequency
\begin{align} \label{omegap2}
\omega_p=\frac{3}{16\pi^2}\frac{\Gamma}{R_\perp^2} \sqrt{f_{xx} f_{yy} } \; .
\end{align}
This is in perfect agreement with the results recently found in~\cite{Kevrekidis:2017aa} by more traditional methods.

Notice that for the harmonic potential \eqref{eq:ellip}, the typical transverse size is $R_\perp \sim c_s/ \omega_x \sim c_s /\omega_y$ (assuming $\omega_x \sim \omega_y$), so Eqs.~\eqref{eq:omegap} and \eqref{omegap2} scale in the same way with $R_\perp$ and $\Gamma$, but the harmonic case \eqref{eq:omegap} has a logarithmic enhancement that the anharmonic case lacks. 

Notice also that the $f_{ab}$ tensor defined in Eq.~\eqref{finite integral} depends not only on the specific function $f$ that defines the trapping potential, Eq.~\eqref{parametrize}, but also on how the integral in Eq.~\eqref{finite integral} is cut off at  $\vec x_\perp \sim R_\perp$. At the edge of the cloud, our perturbative approximations  break down, so we cannot predict at present what is the most physical way to implement this cutoff. For the time being, we therefore leave $f_{ab}$ undetermined; it is plausible that RG ideas like those that led us to Eq.~\eqref{RG n} might help us understand how an integral like Eq.~\eqref{finite integral} is made finite in the IR.


\section{Discussion} \label{discussion}

We close with some comments on our results and possible generalizations.

First, notice that our effective action for the vortex line, Eq.~\eqref{eq:SNR}, is valid for {\em any} trapping potential. This means that our result can also be applied to homogeneous superfluids ``in a box," such as those realized in~\cite{Gaunt:2013aa,Chomaz:2015aa,Mukherjee:2017aa}. 
Our analysis in Sec.~\ref{sec:shallow} indicates that even for potentials that are arbitrarily flat near the center, the vortex will still exhibit precession with an angular frequency scaling as $\omega_p \sim \Gamma / R_\perp^2$, which is a well-known result (see {\it e.g.}~\cite{saffman1992vortex,schwarz}). Moreover, the orbits near the center are always elliptical (or circular), regardless of the shape of the cloud.

Second, notice that the effective string potential \eqref{V eff}  is {\em negative} definite (on general grounds the coupling $T_{(01)}$ is  expected to be positive \cite{Horn:2015zna}). This can lead to instabilities once the interactions of the vortex with phonons are taken into account; for instance, at non-zero temperature we expect the phonon thermal bath to create an effective friction for the precessing vortex, making it slowly migrate to regions of lower and lower effective potential, that is, away from the center of the cloud---see {\it e.g.} \cite{lundh}. It would be interesting to use our EFT to quantify the effect. In particular,
the inclusion of finite temperature effects should not require too much effort.

Lastly, one could generalize our results to systems with trapping along the $z$-direction as well. 
There are experimental indications that vortices in that case get bent by the trap~\cite{Ku:2014aa}, and it would be interesting to see how that effect arises within our EFT.


\begin{acknowledgements}
We are grateful to A.~Morales for introducing us to this interesting phenomenon, and to G.~Iwata, R.~McNally, and K.~Wenz for useful discussions on trapping techniques. We also thank E.~A.~Cornell, S.~Endlich,  A.~Pilloni, M.~Zwierlein, and especially R.~Carretero, P.~Kevrekidis, R.~Penco, and S.~Will for enlightening discussions. This work has been supported by the US Department of Energy grant DE-SC0011941.
\end{acknowledgements}


\appendix*
\section{Relativistic corrections} \label{app}

\begin{figure}[t]
\includegraphics[scale=0.5]{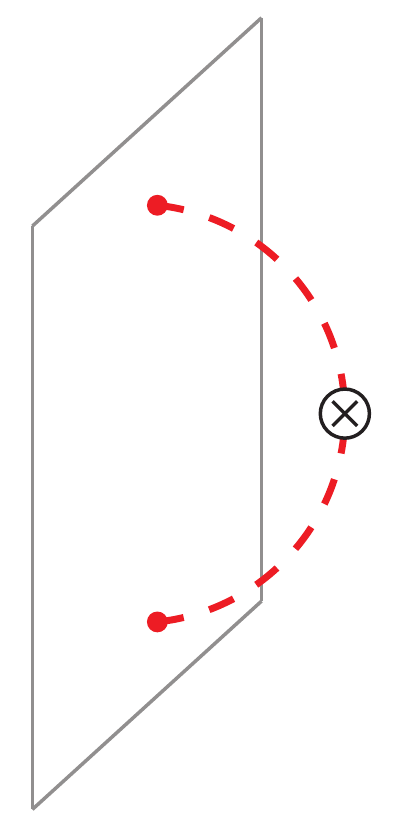}
\caption{Leading relativistic vortex/trap interaction. The red dashed line is again a hydrophoton propagator, and the crossed circle an interaction with the trapping potential.}  \label{fig:reldiagrams}
\end{figure}

Using the  EFT methods outlined in the main text,
with a minimal amount of extra work we can compute the relativistic corrections to the results of Sect.~\ref{2D}. Neglecting terms involving time derivatives and the string's velocity for the same reasons as before, the only new terms we should consider in the action read (see Eqs.~\eqref{eq:SI} and \eqref{rhoij})
\begin{align}
S \supset  \int d^3x dt & \, \frac{\bar n}{2c^2} U_{ij}(\xp)\big(\vec\nabla\times\vec A\,\big)^i\big(\vec\nabla\times\vec A\,\big)^j\; ,
\end{align}
with
\begin{equation} \label{Uij}
U_{ij} (\xp) \equiv V(\xp ) \delta_{ij} - V_{ij}(\xp)  \; ,
\end{equation}
and which, combined with the source \eqref{eq:JBk} for $A$, can give a vortex/trapping potential interaction mediated by $A$ as in the diagram  of Fig.~\ref{fig:reldiagrams}.

After straightforward algebra, the new contribution is found to be
\begin{widetext}
\begin{align} \label{eq:SVA}
   S_{\text{eff}}[\vec X] & \supset -\frac{\bar n^3\lambda^2c^2}{2\bar w^2}\int d^3 xdt\,\epsilon^{ab}\epsilon^{cd}U_{ac}(\xp+\X)\int\frac{d^2 p_\perp d^2 q_\perp}{(2\pi)^4}e^{-i(\vec{p}_\perp+\vec{q}_\perp)\cdot\xp}\frac{p^b_\perp\,q^d_\perp}{p_\perp^{\,2} q_\perp^{\,2}} \notag \\
   &=\frac{\bar n^3\lambda^2 c^2}{8\pi^2\bar w^2}\int dt d\sigma \int d^2 x_\perp\,\epsilon^{ab}\epsilon^{cd}U_{ac}(\xp+\X)\frac{x_\perp^b\,x_\perp^d}{x_\perp^4} \; .
\end{align}
\end{widetext}
Together with Eq.~\eqref{eq:SNR} and replacing $m\to\bar w/\bar nc^2$, this new term gives the complete relativistic vortex action
\begin{widetext}
\begin{align}
  S_{\text{eff}} [\vec X] &= \int dtd\sigma\Big[\frac{\bar n\lambda}{3}\epsilon_{ab}X^a\dot X^b+\frac{2T_{(01)}\bar n c^2}{\bar w c_s^2}V(\X)+\frac{\bar n^3\lambda^2c^4}{8\pi^2\bar w^2c_s^2}\int d^2x_\perp\frac{V(\vec x_\perp+\vec X)}{x_\perp^2} \\
&  \quad-\frac{\bar n^3\lambda^2c^2}{8\pi^2\bar w^2}\int d^2x_\perp \epsilon^{ab}\epsilon^{cd}V_{ac}(\vec x_\perp+\X)\frac{x_\perp^bx_\perp^d}{x_\perp^4}\Big].\notag
\end{align}
\end{widetext}
It is interesting to note that, in the absence of coupling of the trapping to the superfluid velocity, {\it i.e.} for $V_{ab}=0$, the relativistic result is formally equal to the one in Eq.~\eqref{eq:SNR}, since the relativistic correction in Eq.~\eqref{eq:SA} cancels exactly. The same considerations of Sec.~\ref{2D} apply to this action.


\bibliographystyle{apsrev4-1}
\bibliography{biblio}

\end{document}